\documentclass[ba]{imsart2}\usepackage[]{graphicx}\usepackage[]{color}
\makeatletter
\def\maxwidth{ %
  \ifdim\Gin@nat@width>\linewidth
    \linewidth
  \else
    \Gin@nat@width
  \fi
}
\makeatother

\definecolor{fgcolor}{rgb}{0.345, 0.345, 0.345}

\usepackage{framed}
\makeatletter
 {\par\unskip\endMakeFramed%
 \at@end@of@kframe}
\makeatother

\definecolor{shadecolor}{rgb}{.97, .97, .97}
\definecolor{messagecolor}{rgb}{0, 0, 0}
\definecolor{warningcolor}{rgb}{1, 0, 1}
\definecolor{errorcolor}{rgb}{1, 0, 0}
\newenvironment{knitrout}{}{} % an empty environment to be redefined in TeX

\usepackage{alltt}\usepackage[]{graphicx}\usepackage[]{color}
\makeatletter
\def\maxwidth{ %
  \ifdim\Gin@nat@width>\linewidth
    \linewidth
  \else
    \Gin@nat@width
  \fi
}
\makeatother
\usepackage{amsthm}
\usepackage{amsmath}
\usepackage{natbib}
\usepackage[colorlinks,citecolor=blue,urlcolor=blue,filecolor=blue,backref=page]{hyperref}
\usepackage{graphicx}
\usepackage[utf8]{inputenc}
\startlocaldefs
\usepackage{url} 
\usepackage{booktabs} 
\usepackage{color}
\usepackage{todonotes}
\usepackage[printonlyused]{acronym}
\usepackage[toc, page]{appendix}
\usepackage{xifthen}            % for \isempty
\usepackage{array}
\usepackage{amssymb}

\newcommand{\blanco}[1]{  } 
\newcommand{\latin}[1]{\textit{#1}}

\DeclareRobustCommand\xdot{\futurelet\token\Xdot}
\def\Xdot{%
  \ifx\token\bgroup.%
  \else\ifx\token\egroup.%
  \else\ifx\token\/.%
  \else\ifx\token\ .%
  \else\ifx\token!.%
  \else\ifx\token,.%
  \else\ifx\token:.%
  \else\ifx\token;.%
  \else\ifx\token?.%
  \else\ifx\token/.%
  \else\ifx\token'.%
  \else\ifx\token).%
  \else\ifx\token-.%
  \else\ifx\token+.%
  \else\ifx\token~.%
  \else\ifx\token.%
  \else.\ %
  \fi\fi\fi\fi\fi\fi\fi\fi\fi\fi\fi\fi\fi\fi\fi\fi%
}

\newcommand{\eg}{\abk{\latin{e.\,g}}}
\newcommand{\ie}{\abk{\latin{i.\,e}}}

\newlength{\halbebreite}
\DeclareMathOperator{\Nor}{N} % Normal -
\renewcommand{\P}{\operatorname{\mathsf{Pr}}} % Wahrscheinlichkeitsmaß
\newcommand{\p}{f}%{\operatorname{{p}}} % Density function
\DeclareMathOperator{\BF}{BF} % Bayes factor
\newcommand{\partialv}[3][1]{%
\ifthenelse{#1 = 1}{\frac{\partial #2}{\partial #3}}{\frac{\partial^{#1} #2}{\partial #3^{#1}}}
} 

\newcommand{\partials}[3][1]{%
\ifthenelse{#1 = 1}{\frac{d #2}{d #3}}{\frac{d^{#1} #2}{d #3^{#1}}}
} 

\newcommand{\dseps}[2][1]{%
\ifthenelse{#1 = 1}{\frac{d}{d #2}}{\frac{d^{#1}}{d #2^{#1}}}
}

\newcommand{\dsepv}[2][1]{%
\ifthenelse{#1 = 1}{\frac{\partial}{\partial #2}}{\frac{\partial^{#1}}{\partial #2^{#1}}}
}

\newcommand{\ml}[2][1]{% % für Maximum-Likelihood-Schätzer von #1
\ifthenelse{#1 = 1}%
 {\hat{#2}_{\scriptscriptstyle{\mathrm{ML}}}}% 
 {\hat{#2}^{#1}_{\scriptscriptstyle{\mathrm{ML}}}}% z.B. für sigmadach^2
}
\newcommand{\map}[2][0]{% % für MAP-Schätzer von #1
\ifthenelse{#1 = 0}%
 {\hat{#2}_{\scriptscriptstyle{\mathrm{MAP}}}}% 
 {\hat{#2}_{{\scriptscriptstyle{\mathrm{MAP}}_{#1}}}}% z.B. für sigmadach^2
}
\newcommand{\mpm}[2][0]{% % für MPM-Schätzer von #1
\ifthenelse{#1 = 0}%
 {\hat{#2}_{\scriptscriptstyle{\mathrm{MPM}}}}% 
 {\hat{#2}_{{\scriptscriptstyle{\mathrm{MPM}}_{#1}}}}% z.B. für sigmadach^2
}

\newcommand{\given}{\,\vert\,} % für "X gegeben Y" also $X\given Y$ schreiben
\newcommand{\abs}[1]{\left\lvert#1\right\rvert} % Absolutbetrag
\renewcommand{\eg}{\latin{e.\,g.~}}
\renewcommand{\ie}{\latin{i.\,e.~}}

\newcommand{\FPR}{\mbox{FPR}}

\newcommand{\data}{\mbox{data}}

\newcommand{\prep}{p_{\scriptsize \mbox{rep}}}
\newcommand{\pIC}{p_{\scriptsize \mbox{IC}}}
\acrodef{FPR}{false positive risk}
\acrodef{FPRP}{false positive report probability}
\acrodef{minBF}{minimum Bayes factor}
\endlocaldefs

\IfFileExists{upquote.sty}{\usepackage{upquote}}{}
\begin{document}

%% *** Frontmatter *** 
% =============================================================================
\begin{frontmatter}
% \title{Challenging Statistical Significance: \\ A Case for Reverse-Bayes}
% \title{Reverse-Bayes inference: a review of recent advances and their 
% implications for research workers}
%% \title{Reverse-Bayes methods: Recent technical advances with application to meta-analysis}
%% \title{Reverse-Bayes methods: \\ A review of recent technical advances}
\title{Reverse-Bayes methods for evidence assessment and research synthesis}
\runtitle{Reverse-Bayes methods}

\begin{aug}
\author{\fnms{Leonhard} \snm{Held}\thanksref{addr1,t1,t2}
        \ead[label=e1]{leonhard.held@uzh.ch}},
\author{\fnms{Robert} \snm{Matthews}\thanksref{addr2,t1}
        \ead[label=e2]{rajm@physics.org}},
\author{\fnms{Manuela} \snm{Ott}\thanksref{addr1,addr3} 
       \ead[label=e3]{manuela.ott@snf.ch}},
\and
\author{\fnms{Samuel} \snm{Pawel}\thanksref{addr1,t1} 
       \ead[label=e4]{samuel.pawel@uzh.ch}}

\runauthor{L. Held et al.}

\address[addr1]{Department of Biostatistics, University of Zurich,
\printead{e1}, \printead{e4}
}

\address[addr2]{Department of Mathematics, Aston University,
\printead{e2}
}

\address[addr3]{Data Team, Swiss National Science Foundation, 
\printead{e3}
}

\thankstext{t1}{Supported by the Swiss National Science
Foundation (\url{http://p3.snf.ch/Project-189295})}
% \thankstext{t3}{All analyses were performed in the R programming language version 
% paste(version$major, version$minor, sep = ".") \citep{R}. 
% The code to reproduce this manuscript is available at 
% \url{https://gitlab.uzh.ch/}.}

\end{aug}

\begin{abstract}
  It is now widely accepted that the standard inferential toolkit used
  by the scientific research community -- null-hypothesis significance
  testing (NHST) -- is not fit for purpose. Yet despite the threat
  posed to the scientific enterprise, there is no agreement concerning
  alternative approaches for evidence assessment. 
  This lack of consensus reflects
  long-standing issues concerning Bayesian methods, the principal
  alternative to NHST. We report on recent work that builds on an
  approach to inference put forward over 70 years ago 
  to address the
  well-known ``Problem of Priors'' in Bayesian analysis, by reversing
  the conventional prior-likelihood-posterior (``forward'') use of
  Bayes's Theorem. Such Reverse-Bayes analysis allows priors to be
  deduced from the likelihood by requiring that the posterior achieve
  a specified level of credibility. We summarise the technical
  underpinning of this approach, and show how it opens up new
  approaches to common inferential challenges, 
  such as assessing the credibility of scientific findings, setting
  them in appropriate context, estimating the probability of
  successful replications, and extracting more insight from NHST while
  reducing the risk of misinterpretation.  We argue that Reverse-Bayes
  methods have a key role to play in making Bayesian methods more
  accessible and attractive for evidence assessment and research synthesis. 
  As a running
  example we consider a recently published meta-analysis from several
  randomized controlled clinical trials investigating the association
  between corticosteroids and mortality in hospitalized patients with
  COVID-19.
\end{abstract}

%% ** Keywords **

\end{frontmatter}

%% ** Mainmatter **
% =============================================================================
\section{Introduction: the origin of Reverse-Bayes methods}\label{sec:intro}
\begin{center}
\begin{minipage}{10cm}
  { { { ``We can make judgments of initial probabilities and infer
        final ones, or we can equally make judgments of final ones and
        infer initial ones by \emph{Bayes's theorem in reverse}.'' }}}
%\vspace{.5cm}
\end{minipage}
\end{center}
\begin{flushright}
\citet[p.~29]{good:1983}
\end{flushright}

There is now a common consensus that the most widely-used methods of 
statistical inference have led to a crisis in both the interpretation of
research findings and their replication \citep[\eg][]{Gelman2014, Wasserstein2016}.  
At the same time, there is a lack of consensus on how to address the challenge \citep{Matthews:2017}, 
as highlighted by the plethora of alternative techniques to null-hypothesis
significance testing now being put forward \citep[see \eg][and references therein]{Wasserstein2019}.
% and references therein). 
Especially striking is the relative dearth of 
alternatives based on Bayesian concepts. Given their intuitive inferential
basis and output \citep[see \eg][or some
other textbook]{Wagenmakers2008, McElreath2018}, these would  seem obvious 
candidates to supplant the prevailing frequentist methodology. However, 
it is well-known that the adoption of Bayesian methods continues to be hampered
by several factors, such as the belief that advanced computational tools are 
required to make Bayesian statistics practical \citep[\eg][]{Green2015}.
The most persistent of these is that the full benefit of Bayesian 
methods demands specification of a prior level of belief, even in the absence
of any appropriate insight. This ``Problem of Priors'' has cast a shadow over
Bayesian methods since their emergence over 250 years ago 
\citep[see \eg][]{McGrayne2011}, and has led to a variety of approaches, such as 
prior elicitation, prior sensitivity analysis, and objective Bayesian methodology;
all have their supporters and critics. 

One of the least well-known was suggested over 70 years ago
\citep{good:1950} by one of the best-known proponents of Bayesian
methods during the 20\textsuperscript{th} century, I.J.~Good.  It
involves reversing the conventional direction of Bayes's Theorem and
determining the level of prior belief required to reach a specified
level of posterior belief, given the evidence observed. This reversal
of Bayes's Theorem allows the assessment of new findings on the basis
of whether the resulting prior is reasonable in the light of existing
knowledge.  Whether a prior is plausible in the light of existing
knowledge can be assessed informally or more formally using techniques
for comparing priors with existing data as suggested by
\citet{box:1980} and further refined by \citet{evans.moshonov2006}.
%% If this is
%% the case, then Bayes's Theorem implies the strength of the new evidence reaches
%% the standard set by the pre-specified posterior level of belief.\todo{MO: This sentence was hard to understand for me at first. Perhaps start with "If the prior is plausible, then" instead of "If this is the case"?}
Good stressed 
that despite the routine use of the adjectives ``prior'' and ``posterior'' in
applications of Bayes's Theorem, the validity of any resulting inference does not
require a specific temporal ordering, as the theorem is simply a constraint 
ensuring consistency with the axioms of probability. While reversing Bayes's Theorem
is still regarded as unacceptable by some on the grounds it allows ``cheating''
in the sense of choosing priors to achieve a desired posterior inference
\citep[\eg][p. 143]{Ohagan2004}, others point out this is not an ineluctable
consequence of the reversal \citep[\eg][p. 78--79]{cox:2006}. As we shall
show, recent technical advances further weaken this criticism.

% Yet despite Good's belief that Reverse-Bayes methods ``will be found to be of 
% the utmost value in future Bayesian statistics'' \citep[p. 29]{good:1983}, 
% their potential has remained largely unexplored. 
Good's belief in the value of Reverse-Bayes methods won support from E.T.~Jaynes
in his well-known treatise on probability. Explaining a specific manifestation
of the approach (to be discussed shortly) Jaynes remarked: ``We shall find it 
helpful in many cases where our prior information seems at first too vague to 
lead to any definite prior probabilities; it stimulates our thinking and tells
us how to assign them after all'' \citep[p. 126]{Jaynes2003}. Yet despite the 
advocacy of two leading figures in the foundations of Bayesian methodology,
the potential of Reverse-Bayes methods has remained largely unexplored. 
Most published work has focused 
on their use in putting new research claims in context, with Reverse-Bayes 
methods being used to assess whether the prior evidence needed to make a claim 
credible is consistent with existing insight 
\citep{Carlin1996,matthews:2001,matthews:2001b,spiegelhalter2004,greenland:2006,greenland:2011,held2013,Colquhoun:2017,Colquhoun:2019,held2019,Held2020,Pawel2020b}.

The purpose of this paper is to highlight recent technical
developments of Good's basic idea which lead to inferential tools of
practical value in the analysis of summary measures as reported
in meta-analysis. As a running
example we consider a recently published meta-analysis investigating
the association between corticosteroids and mortality in hospitalized
patients with COVID-19.  Specifically, we show how Reverse-Bayes
methods address the current concerns about the interpretation of new
findings and their replication. We begin by illustrating the basics of
the Reverse-Bayes approach for both hypothesis testing and parameter
estimation. This is followed by a discussion of Reverse-Bayes methods
for assessing effect estimates in Section \ref{sec:effects}.  These
allow the credibility of both new and existing research findings
reported in terms of NHST to be evaluated in the context of existing
knowledge. This enables researchers to go beyond the standard
dichotomy of statistical significance/non-significance, extracting
further insight from their findings. We then discuss the use of the
Reverse-Bayes approach in the most recalcitrant form of the Problem of
Priors, involving the assessment of research findings which are
unprecedented and thus lacking any clear source of prior support. We
show how the concept of intrinsic credibility resolves this challenge,
and puts recent calls to tighten $p$-value thresholds on a principled
basis \citep{Benjamin2017}.  In Section \ref{sec:bfs} we describe
Reverse-Bayes methods with Bayes factors, the principled solution for
Bayesian hypothesis testing. Finally, we describe in Section
\ref{sec:p.equals} Reverse-Bayes approaches to interpretational
issues that arise in conventional statistical analysis based on
$p$-values, and how they can be used to flag the risk of inferential
fallacies.  We close with some extensions and final conclusions.

\subsection{Reverse-Bayes for hypothesis testing}
The subjectivity involved in the specification of prior distributions
is often seen as a weak point of Bayesian inference. The Reverse-Bayes
approach can help to resolve this issue both in hypothesis testing and
parameter estimation, we will start with the former.

Consider a null hypothesis $H_0$ with prior
probability $\pi=\P(H_0)$, so $\P(H_1) =1-\pi$ is the prior
probability of the alternative hypothesis $H_1$.
Computation of the posterior probability of $H_0$ is routine with Bayes' theorem: 
\[
  \Pr(H_{1} \given \data) = \frac{\Pr(\data \given H_{1}) \Pr(H_{1})}{\Pr(\data \given H_{0}) \Pr(H_{0}) + 
    \Pr(\data \given H_{1}) \Pr(H_{1})}.
\]
Bayes' theorem can be written in more compact form as
\begin{equation}\label{eq:eq0}
  \frac{\Pr(H_{1} \given \data)}{\Pr(H_{0} \given \data)} =
  \frac{\Pr(\data \given H_{1})}{\Pr(\data \given H_{0})} \,  \frac{\Pr(H_{1})}{\Pr(H_{0})}, 
\end{equation}
\ie the posterior odds are the likelihood ratio times the prior odds. The
standard 'forward-Bayes' approach thus fixes the prior odds (or one of the underlying 
probabilities), determines the likelihood ratio for the available data, 
and takes the product to compute the posterior
odds. Of course, the latter can be easily back-transformed to the posterior
probability $\Pr(H_{1} \given \data)$, if required. 
The Problem of Priors is now apparent: in order for us to update the odds in 
favour of $H_1$, we must first specify the prior odds. This can be 
problematic in situations where, for example, the evidence on which to base 
the prior odds is controversial or even non-existent. 

However, as Good emphasised it is entirely justifiable to ``flip'' Bayes's 
theorem around, allowing us to ask the question: Which prior, when combined 
with the data, leads to our specified posterior?
\begin{equation}\label{eq:eq1}
  \frac{\Pr(H_{1})}{\Pr(H_{0})} = \left. \frac{\Pr(H_{1} \given \data)}{\Pr(H_{0} \given \data)} \, \Big/ 
\,   \frac{\Pr(\data \given H_{1})}{\Pr(\data \given H_{0})} \right. .
\end{equation}
%% A third way is to fix prior and posterior odds and solve for the
%% likelihood ratio.  In what follows we will concentrate in this paper
%% on the reverse-Bayes approach.

For illustration we re-visit an example put forward by
\citet[p.~35]{good:1950}, perhaps the first published Reverse-Bayes
calculation. 
It centres on a question for which the setting of an initial prior is 
especially problematic: does an experiment provide convincing evidence for
the existence of extra-sensory perception (ESP)?
The substantive hypothesis $H_{1}$ is that ESP exists, so that $H_{0}$
asserts it does not exist. 
Imagine an experiment in which a person has to make $n$ consecutive guesses of 
random digits (between 0 and 9)
and all are correct. The likelihood ratio is therefore
\[
   \frac{\Pr(\data \given H_{1})}{\Pr(\data \given H_{0})} = \frac{1}{(1/10)^n} = 10^n .
\]
It is unlikely that sceptics and advocates of the existence of ESP would ever 
agree on what constitutes reasonable priors from which to start a standard 
Bayesian analysis of the evidence. However, Good argued that Reverse-Bayes 
offers a way forward by using it to set bounds on the prior probabilities for
$H_{1}$ and $H_{0}$. This is achieved via the outcome of an imaginary 
(Gedanken) experiment capable of demonstrating $H_{1}$ is more likely than 
$H_{0}$, that is, of leading to posterior probabilities such that
$\Pr(H_1 \given \data) > \Pr(H_0 \given \data)$. Using this approach, which 
Good termed the \emph{Device of Imaginary Results}, we see that if the ESP 
experiment produced 20 correct consecutive guesses, \eqref{eq:eq1} implies 
that ESP may be deemed more likely than not to exist by anyone whose priors 
satisfy ${\Pr(H_{1})}/{\Pr(H_{0})} > 10^{-20}$. In contrast, if only $n = 3$
correct guesses emerged, then the existence of ESP could be rejected by anyone 
whose priors satisfy ${\Pr(H_{1})}/{\Pr(H_{0})} < 10^{-3}$. Using Bayes's 
Theorem in reverse has thus led to a quantitative statement of the prior beliefs
that either advocates or sceptics of ESP must be able to justify in the face of
results from a real experiment. 
The practical value of Good's approach was noted by Jaynes in his treatise:
``[I]n the present state of development of probability theory, the device of 
imaginary results is usable and useful in a very wide variety of situations,
where we might not at first think it applicable'' 
\citep[p. 125--126]{Jaynes2003}.
     
It is straightforward to extend \eqref{eq:eq0} and \eqref{eq:eq1} to hypotheses
that involve unknown parameters $\theta$. The likelihood ratio 
$\Pr(\data \given H_{1})/\Pr(\data \given H_{0})$ is then called a Bayes
factor \citep{jeffreys:1961, kass1995} where 
\[
 \Pr(\data \given H_{i}) = \int \Pr(\data \given \theta, H_{i}) \p(\theta \given H_{i}) d \theta %% \quad i=0,1,
\]
is the marginal likelihood under hypothesis $H_i$, $i=0,1$,
obtained be integration of the ordinary likelihood with respect to the
prior distribution $\p(\theta \given H_i)$. We will apply the Reverse-Bayes approach to
Bayes factors in Section \ref{sec:bfs} and \ref{sec:p.equals}.

\subsection{Reverse-Bayes for parameter estimation}
% If the likelihood function stems from a continuous probability
% distribution, then the likelihood ratio is
% ${\p(\data \given H_{1})}/{\p(\data \given H_{0})}$, here
% $\p(\data \given H_i)$ denotes the probability density function of the
% data under the two hypotheses $H_i$, $i=0,1$.
We can also apply the Reverse-Bayes idea to continuous prior and posterior
distributions of a parameter of interest $\theta$. Reversing Bayes' theorem 
\[
  \p(\theta \given \data) = \frac{\p(\data \given \theta) \p(\theta)}{\p(\data)}
  %% {\int \p(\data \given \theta) \p(\theta) d\theta}.
\]
then leads to 
\begin{equation}
  \p(\theta) = \p(\data) \, \frac{\p(\theta \given \data)}{\p(\data \given \theta)}.
  %% {\int \p(\data \given \theta) \p(\theta) d\theta}.
\label{eq:revBayesDensity}
\end{equation}
So the prior is proportional to the posterior divided by the likelihood with 
proportionality constant $\p(\data)$.

Consider Bayesian inference for the mean $\theta$ of a univariate
normal distribution, assuming the variance $\sigma^2$ is known.
Let $x$ denote the observed value from that $\Nor(\theta, \sigma^2)$ 
distribution and suppose the prior for $\theta$ (and hence also the posterior) 
is normal.  Each of them is determined by two parameters, usually
the mean and the variance, but two distinct quantiles would also
work. If we fix both parameters of the posterior, then the prior in
\eqref{eq:revBayesDensity} is -- under a certain regularity condition --
uniquely determined. For ease of presentation we work with
the observational precision $\kappa=1/\sigma^2$ and denote the prior
and posterior precision by
$\delta$ and $\delta'$, respectively.  Finally let $\mu$ and
$\mu'$ denote the prior and posterior mean, respectively.

Forward-Bayesian updating tells us how to compute the posterior
precision and mean: 
\begin{eqnarray}
\label{eq:posterior-precision}
\delta' & = & \delta + \kappa, \\
\mu' & = & \frac{\mu \delta + x \kappa}{\delta'}.\label{eq:posterior-mean}
\end{eqnarray}
For example, fixed-effect (FE) meta-analysis is based on iteratively
applying \eqref{eq:posterior-precision} and \eqref{eq:posterior-mean}
to the summary effect estimate $x_i$ with standard error $\sigma_i$
from the $i$-th study, $i=1,\ldots,n$, starting with an initial 
precision of zero.
Reverse-Bayes simply inverts these equations, which leads to the following:
\begin{eqnarray}
\label{eq:prior-precision}
\delta & = & \delta' - \kappa,  \\
\mu & = & \frac{\mu'\delta' - x \kappa}{\delta}, \label{eq:prior-mean}
\end{eqnarray}
provided $\delta' > \kappa$, \ie the posterior 
precision must be larger than the observational precision.

\begin{figure}[!htb]
\begin{knitrout}
\definecolor{shadecolor}{rgb}{0.969, 0.969, 0.969}\color{fgcolor}

{\centering \includegraphics[width=\linewidth]{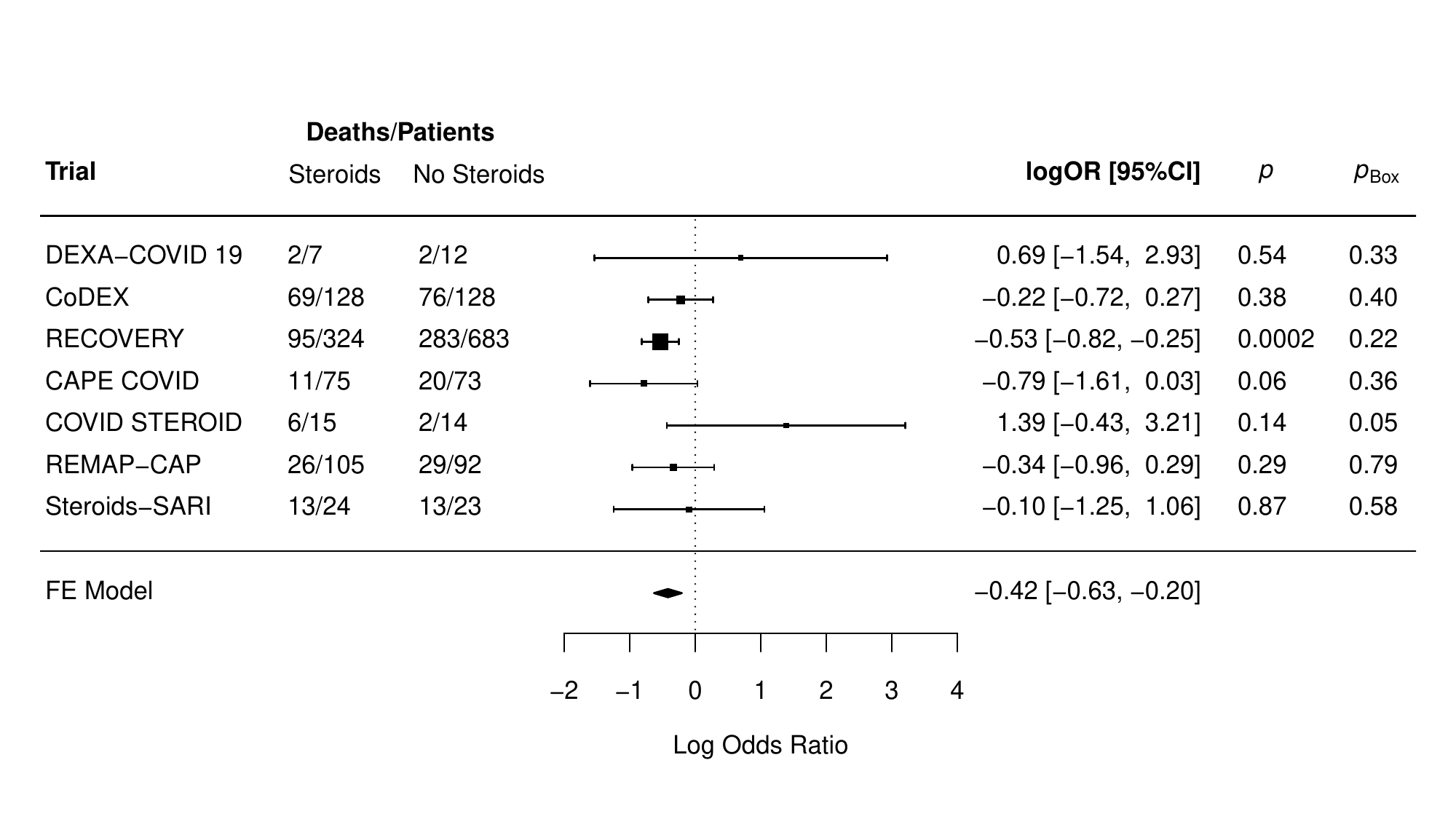} 

}

\end{knitrout}
\caption{Forest plot of fixed-effect meta-analysis of randomized clinical
trials investigating association between corticosteroids and mortality in
hospitalized patients with COVID-19 \citep{REACT2020}. Shown are number of deaths
among total number of patients for treatment/control group, log odds ratio 
effect estimates with 95\% confidence interval, two-sided $p$-values $p$, and 
prior-predictive tail probabilities $p_{\text{Box}}$ with a meta-analytic estimate
based on the remaining studies serving as the prior.}
\label{fig:covid19-meta}
\end{figure}

We will illustrate the application of \eqref{eq:prior-precision} and
\eqref{eq:prior-mean} as well as the methodology in the rest of this
paper using a recent meta-analysis combining information from $n=7$
randomized controlled clinical trials investigating the association
between corticosteroids and mortality in hospitalized patients with
COVID-19 \citep{REACT2020}; its results are reproduced in Figure
\ref{fig:covid19-meta} (here and henceforth, odds ratios (ORs) are
expressed as log odds ratios to transform the range from $(0, \infty)$
to $(-\infty, +\infty)$, consistent with the assumption of normality).
Let $x_i = \hat \theta_i$ denote the maximum likelihood estimate (MLE)
of the log odds ratio $\theta$ in the $i$-th study with standard error
$\sigma_i$.  The meta-analytic odds ratio estimate under the
fixed-effect model (the pre-specified primary analysis) is
$\widehat{\text{OR}} = 0.66$ [95\% CI,
0.53, 0.82], respectively
$\hat \theta = -0.42$ [95\% CI,
-0.63, -0.20] for the log odds ratio $\theta$,
indicating evidence for lower mortality of patients treated with
corticosteroids compared to patients receiving usual care or placebo.
The pooled effect estimate $\hat \theta$ represents a posterior mean
$\mu'$ with posterior precision $\delta'=83.8$.

With a meta-analysis such as this, it is of interest to quantify
potential conflict among the effect estimates from the different
studies.  To do this, we follow \citet{Presanis2013} and compute a
prior-predictive tail probability \citep{box:1980, evans.moshonov2006}
for each study-specific estimate $\hat \theta_i$, with a meta-analytic
estimate based on the remaining studies serving as the prior. As
discussed above, fixed-effect meta-analysis is standard
forward-Bayesian updating for normally distributed effect estimates
with an initial flat prior.  Hence, instead of fitting a reduced
meta-analysis for each study, we can simply use the the Reverse-Bayes
equations \eqref{eq:prior-precision} and \eqref{eq:prior-mean}
together with the overall estimate to compute the parameters of the
prior in the absence of the $i$-th study (denoted by the index $-i$):
\begin{eqnarray*}
\delta_{-i} & = & \delta' - 1/\sigma_i^2, \\
\mu_{-i} & = & \frac{ \mu' \delta' - \hat \theta_i / \sigma_i^2}{\delta_{-i}}.
\end{eqnarray*}
For example, through omitting the
\citet{RECOVERY2020}  trial result $\hat{\theta}_i =  -0.53$ with
standard error
${\sigma}_i =  0.145$  we obtain 
$\delta_{-i} = 36.1$ and $\mu_{-i} = -0.26$. 
%% \begin{eqnarray*}
%% \delta_{-i} & = & round(kp, 1) - 1/round(sqrt(sigma2), 3)^2=round(k0, 1), \\
%% \mu_{-i} & = &  \frac{round(mup, 3) \cdot round(kp, 1)- (round(theta, 3)) / round(sqrt(sigma2), 3)^2}{round(k0, 1)} =round(mu0, 2).
%% \end{eqnarray*}
%% the prior mean 
%% $\mu_{-i}=round(mu0, 2)$ and 95\% prior %% highest prior density
%% credible interval from paste(round(mu0 + c(-1, 1)*qnorm(p = 0.975)*sqrt(1/k0), 2), collapse = " to ").  
A prior predictive tail probability using the approach from
\citet{box:1980} is then obtained by computing
$p_{\text{Box}} = \P(\chi^2_1 \geq t^2_{\text{Box}})$ with
$$
t_{\text{Box}} = \frac{\hat \theta_i - \mu_{-i}}{\sqrt{\sigma_i^2 +
    1/\delta_{-i}}} =-1.24.$$ This
leads to $p_{\text{Box}} = 0.22$ for
the RECOVERY trial, indicating very little prior-data conflict.  The
tail probabilities for the other studies are even larger, with the
exception of the COVID STEROID trial ($p_{\text{Box}}=0.05$),
see Figure \ref{fig:covid19-meta}. The lack of strong conflict can be
seen as an informal justification of the assumptions of the underlying
fixed-effect meta-analysis \citep{Presanis2013,Ferkingstad_etal2017}.
A related method in network meta-analysis is to assess consistency via 
"node-splitting" \citep{dias.etal2010}.

Instead of determining the prior completely based on the posterior, 
one may also want to fix one parameter of the posterior
and one parameter of the prior. This is of particular interest in order to 
challenge ``significant'' or ``non-significant'' findings through the Analysis of
Credibility, as we will see in the following section.

\section{Reverse-Bayes methods for the assessment of effect estimates}\label{sec:effects}
A more general question amenable to Reverse-Bayes methods is the
assessment of effect estimates and their statistical significance or
non-significance. This issue has recently attracted intense interest
following the public statement of the American Statistical Association
about the misuse and misinterpretation of the NHST concepts of
statistical significance and non-significance \citep{Wasserstein2016}.
First investigated 20 years ago in \citet{matthews:2001} with
subsequent discussion in \citet{matthews:2001b}, Reverse-Bayes methods
for assessing both statistically significant and non-significant
findings has been termed the Analysis of Credibility \citep[or
AnCred,][]{matthews2018}, whose principles and practice we now briefly
review.

\subsection{The Analysis of Credibility}
\label{sec:AnCred}
Suppose the study gives rise to a conventional confidence interval for
the unknown effect size $\theta$ at level $1 - \alpha$ with lower limit
$L$ and upper limit $U$.  Assume that $L$ and $U$ are symmetric around
the point estimate $\hat \theta$ (assumed to be normally distributed
with standard error $\sigma$).
AnCred then takes this likelihood and uses a Reverse-Bayes approach to
deduce the prior required in order to 
generate evidence for the existence of an effect, in the form of a posterior 
that excludes no effect. As such, AnCred allows evidence deemed 
\emph{statistically significant}/\emph{non-significant} in the NHST framework to
be assessed for its \emph{credibility} in the Bayesian framework. As the latter
represents $\Pr(H_0 \given \text{data})$ and thus a conditioning on the data 
rather than the null hypothesis, it is inferentially directly relevant to
researchers.  
After a suitable transformation AnCred can be applied to 
% a large number
a large number of commonly used effect measures such
as differences in means, odds ratios, relative risks and correlations
\citep[see the literature of meta-analysis for details about conversion among
effect size scales, \eg][Chapter 11.6]{Hedges2019}.
The inversion of Bayes’s Theorem needed to assess credibility requires 
the form and location of the prior distribution to be specified. This in
turn depends on whether the claim being assessed is statistically significant
or non-significant; we consider each below.

\subsubsection{Challenging statistically significant findings}

A statistically significant finding at level $\alpha$ is characterized by both $L$
and $U$ being either positive or negative. Equivalently
$z^2 > z_{\alpha/2}^2$ is required where $z=\hat \theta/\sigma$
denotes the corresponding test statistic and $z_{\alpha/2}$ the
$(1-\alpha/2)$-quantile of the standard normal distribution. 

For significant findings, the idea is to ask how sceptical we would
have to be not to find the apparent %positive 
effect estimate convincing.  To this end, a ``critical prior interval''
\citep{matthews:2001b} with limits $-S$ and $S$ is derived such that the
corresponding posterior credible interval just includes zero, the value of
no effect. This critical prior interval can then be compared with internal or
external evidence to assess if the finding is credible or not, despite being 
``statistically significant''.
% ``statistical significant''.

More specifically, a reverse Bayes approach is applied to significant
confidence intervals (at level $\alpha$) based on a normally
distributed effect estimate. The prior is a ``sceptical'' mean-zero
normal distribution with variance $\tau^2 = g \cdot \sigma^2$, so the only 
free parameter is the relative prior variance $g = \tau^2/\sigma^2$.  
The posterior is hence also normal and either its lower $\alpha/2$-quantile 
(for positive $\hat{\theta}$) or upper $1 - \alpha/2$-quantile 
(for negative $\hat{\theta}$) is fixed to zero, so just represents
``non-credible''. The sufficiently sceptical prior then has 
relative variance 
\begin{align}
\label{eq:sspv}
  g &= 
  \begin{cases} 
    \dfrac{1}{z^2/z_{\alpha/2}^2 - 1}
    & ~~ \text{if} ~ 
    {z^2} > {z_{\alpha/2}^2} \\ 
    \text{undefined} & ~~ \text{else}
  \end{cases} 
  \end{align}
see \citet[Appendix]{held2019} for a derivation. 
The corresponding \emph{scepticism limit} is
\begin{equation}
  \label{eq:S}
  S = \frac{(U-L)^2}{4 \sqrt{UL}},
\end{equation}
which holds for any value of $\alpha$ provided the effect is
significant at that level.

The left plot in Figure \ref{fig:anCredEx} illustrates the AnCred procedure
for the finding from the RECOVERY trial \citep{RECOVERY2020}, the only 
statistically significant result (at the convention $\alpha=0.05$ level) 
shown in Figure \ref{fig:covid19-meta}. 
The trial found a decrease in COVID-19 mortality for patients  
treated with corticosteroids compared to usual care or placebo
($\hat{\theta} = 
-0.53$ [95\% CI, -0.82, -0.25]). 
The sufficiently sceptical prior has relative variance 
$g = 0.39$, so the sufficiently sceptical 
prior variance needs to be roughly 
2.5 times smaller than the variance of 
the estimate to make the result non-credible. 
The scepticism limit on the log odds ratio scale turns out to be
$S=0.18$,
which corresponds to a critical prior interval with limits 
0.84 and 1.19 on the odds ratio scale. 
Thus sceptics may still reject the RECOVERY trial finding as lacking 
credibility despite its statistical significance if external evidence suggests 
mortality reductions (in terms of odds) are unlikely to exceed 
$1 - 0.84 \approx 16 \%$.

It is also possible to apply the approach to the meta-analytic log odds ratio 
estimate $\hat \theta = -0.42$ [95\% CI,
-0.63, -0.20] from all 7 studies combined. Then
$S= 0.13$, so the meta-analytic estimate can be
considered as non-credible if external evidence suggests that
mortality reductions are unlikely to exceed
$1-\exp(-S)=1 - 0.88 \approx
12 \%$. This illustrates that the
meta-analytic estimate has gained credibility compared to the result
from the RECOVERY study alone, despite the reduction in the effect estimate
($\widehat{\text{OR}} = \exp(\hat \theta) = 0.66$ vs.~0.59 in the RECOVERY study).

\subsubsection{Challenging statistically non-significant findings}
It is also possible to challenge ``non-significant'' findings (\ie those
for which the CI  now includes zero, so $z^2 < z_{\alpha/2}^2$) using a prior
that pushes the posterior towards being credible in the Bayesian sense, with 
posterior credible interval no longer including zero, corresponding to no effect. 

\citet{matthews2018} proposed the ``advocacy prior'' for
this purpose, a normal prior with positive mean $\mu$ and variance $\tau^2$ 
chosen such that the $\alpha/2$-quantile is fixed to zero 
(for positive effect estimates $\hat \theta > 0$). 
He showed that the ``advocacy limit'' AL, 
the $(1-\alpha/2)$-quantile of the advocacy prior is
\begin{equation}\label{eq:AL}
  \mbox{AL} = - \frac{U+L}{2 \, U L} (U-L)^2
\end{equation}
to reach credibility of the corresponding posterior at level $\alpha$.
We show in Appendix \ref{app:app1} that the corresponding relative prior mean $m = \mu / \hat \theta$ 
is
\begin{align}
\label{eq:mu}
  m &= 
  \begin{cases} 
    \dfrac{2}{1 - z^2/z_{\alpha/2}^2}
    & ~~ \text{if} ~ 
    {z^2} < {z_{\alpha/2}^2} \\ 
    \text{undefined} & ~~ \text{else. }
  \end{cases} 
  \end{align}
  
    There are two important properties of the 
advocacy prior. First, %% the prior variance is $\tau^2 = \mu^2/z_{\alpha/2}^2$ and
the coefficient of variation CV is 
$$\mbox{CV} = \tau/\mu = z_{\alpha/2}^{-1}.$$
%% because the prior standard deviation is
%% $\tau = \mbox{AL}/(2 \, z_{\alpha/2})$.  
The advocacy prior $\theta \sim \Nor(\mu, \tau^2=\mu^2 \, \mbox{CV}^2)$ is hence
characterized by a fixed coefficient of variation, so this prior has
equal evidential weight (quantified in terms of
$\mu/\tau=z_{\alpha/2}$) as data which are ``just significant'' at
level $\alpha$.
% Second, \eqref{eq:mu} implies that $m > 2$, so the
% prior guess $\mu$ of the advocate needs to be at least twice as large as the
% estimate $\hat \theta$ to make the "non-significant" finding
% significant.  
Second, the advocacy limit AL defines the family of normal priors capable of 
rendering a ``non-significant'' finding credible at the same level. Such priors 
are summarized by the credible interval ($L_o, U_o$) where $L_o \geq 0$, $U_o \leq \mbox{AL}$.  
Thus when confronted with a ``non-significant'' result -- often, and wrongly, 
interpreted as indicating no effect --  advocates of the existence of an effect 
may still claim the existence of the effect is credible to the same level if 
there exists prior evidence or insight compatible with the credible interval 
($L_o, U_o$). %%  where $L_o \geq 0$, $U_o  \leq \mbox{AL}$. 
If the evidence for an effect is weak (strong), the 
resulting advocacy prior will be broad (narrow), giving advocates of an effect 
more (less) latitude to make their case under terms of AnCred. 
Note that \eqref{eq:AL} and \eqref{eq:mu} also hold for
negative effect estimates, where we fix the $(1-\alpha/2)$-quantile of
the advocacy prior to zero and define the advocacy limit AL as the
$\alpha/2$-quantile of the advocacy prior.

\begin{figure}[!htb]
\begin{knitrout}
\definecolor{shadecolor}{rgb}{0.969, 0.969, 0.969}\color{fgcolor}

{\centering \includegraphics[width=\linewidth]{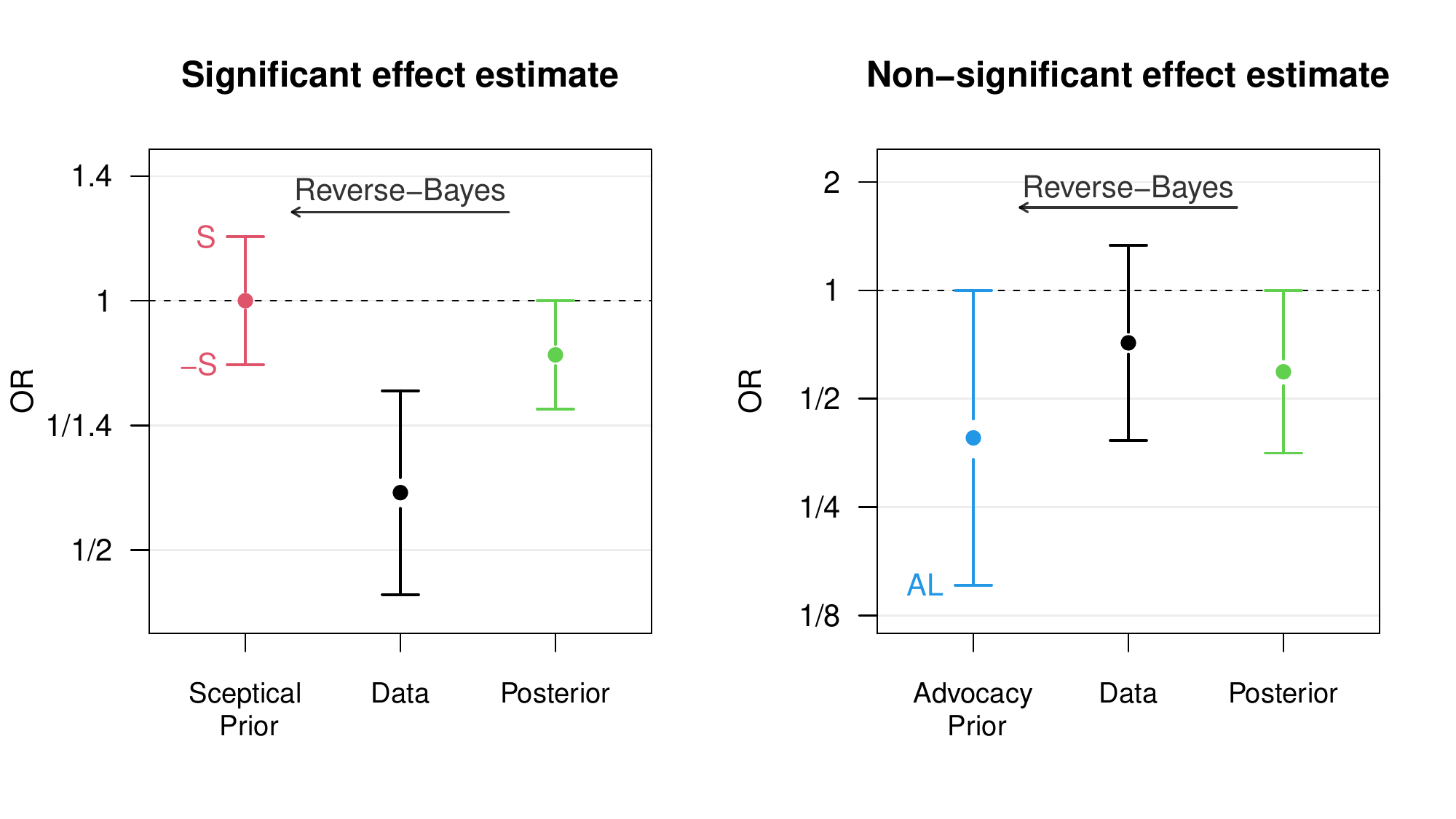} 

}

\end{knitrout}
\caption{Two examples of the Analysis of Credibility. Shown are point estimates 
within 95\% confidence/credible intervals. The left plot
illustrates how a sceptical prior is used to challenge the significant
finding from the RECOVERY trial \citep{RECOVERY2020}. The right plot
illustrates how an advocacy prior is used to challenge a non-significant finding 
from the REMAP-CAP trial \citep{REMAPCAP2020}.  
In both scenarios the posterior is fixed to be just credible/non-credible.}
\label{fig:anCredEx}
\end{figure}

For illustration we consider the data from the REMAP-CAP trial
\citep{REMAPCAP2020}
that supported the RECOVERY trial finding of decreased COVID-19 mortality
from corticosteroid use.
However, this trial involved far fewer patients, and despite the point estimate
showing efficacy, the relatively large uncertainty rendered the overall finding
non-significant at the 5\% level ($\hat{\theta} = -0.34$ 
[95\% CI, $-0.96,  0.29$]).
Such an outcome is frequently (and wrongly) taken to imply no effect. The use 
of AnCred leads to a more nuanced conclusion. 
The advocacy limit AL on the log odds ratio scale for REMAP-CAP is 
$-1.89$, %so $\mu = -0.94$.  
\ie $0.15$ on the odds ratio scale,
see also the right plot in Figure \ref{fig:anCredEx}. 
Thus advocates of the effectiveness of corticosteroids can regard the trial as
providing credible evidence of effectiveness despite its non-significance if
external evidence supports mortality reductions (in terms of odds) in the range
0\% to 85\%.
So broad an advocacy range reflects the fact that this relatively small trial
provides only modest evidential weight, and thus little constraint on prior
beliefs about the effectiveness of corticosteroids.

%% \subsubsection{Data representation of sceptical and advocacy priors}

\subsubsection{Assessing credibility via equivalent prior study sizes}
\citet{greenland:2006} showed that reverse-Bayes credibility
assessments can be formulated in terms of the size of a prior study
capable of challenging a claim of statistical significance.
This perspective helps to put the required weight of prior evidence needed for such a challenge in the context of the observed  data. 
%% has described how to translate priors into the
%% equivalent data from an hypothetical trial that would have produced
%% the assumed prior knowledge. This perspective helps to reveal how
%% priors relate to the actual data and is particularly useful in the
%% application of the Analysis of Credibility. 

For normal priors with
mean $\mu$ and variance $\tau^2$ the equivalent data prior has
$2/\tau^2$ cases in each arm of a trial with a sufficiently large
number of patients in each arm. For example, the sceptical prior for
the RECOVERY trial corresponds to 244 deaths in
each of two arms with, say, 100'000 patients each. If we aim for a
more realistic mortality rate in the hypothetical prior trial, for
example the same as in the RECOVERY trial overall
(37.5\%), then the sceptical prior
corresponds to 389 deaths out of
1038 patients in each arm.  This
is considerably larger than the actual number of deaths in the control
arm (283) and much larger that the number of deaths in the
intervention arm (95) and underlines that the sufficiently sceptical
prior needs to be chosen rather tight to make the RECOVERY trial
result no longer convincing.

On the other hand, the advocacy prior for the REMAP-CAP result
translates into only 9 deaths in each of the two
arms of sufficiently large size, considerably smaller than the
observed number of deaths in the two arms of the study (26
resp.~29). The non-zero prior mean $\mu=-0.94$ can
be incorporated with an allocation ratio of $1:R=\exp(\mu)$,
approximately $5:2$, to shift the prior towards ``advocacy'', for
example with 250'000 patients in the intervention arm and 100'000 in
the control arm.  If we aim for roughly the same control mortality
rate as in the REMAP-CAP trial
(32\%), then the sceptical prior
corresponds to 11 deaths in each arm out of
83 respectively
39 patients.

%% \subsubsection{Relationship between Analysis of Credibility and the fail-safe 
%% $N$ method}
\subsubsection{The fail-safe $N$ method}
%% There is an interesting connection between AnCred and
Another data representation of a sceptical prior forms the basis of
the well-known
``fail-safe $N$'' method, sometimes also called ``file-drawer
analysis''.  This method, first introduced by \citet{Rosenthal1979}
and later refined by \citet{Rosenberg2005}, is commonly applied to the
results from a meta-analysis and answers the question: ``How many
unpublished negative studies do we need to make the meta-analytic
effect estimate non-significant?'' A relatively large $N$ of such
unpublished studies suggests that the estimate is robust to potential
null-findings, for example due to publication bias. Calculations are
made under the assumption that the unpublished studies have an average
effect of zero and a precision equal to the average precision of the published 
ones.

While the method does not identify nor adjust for publication bias, it
provides a quick way to assess %%\todo{MO: "way to assess" instead?} 
how robust the meta-analytic effect
estimate is. The method is available in common software packages such
as \texttt{metafor} \citep{Viechtbauer2010} and its simplicity and
intuitive appeal have made it very popular among researchers.

AnCred and the fail-safe $N$ are both based on the idea to challenge
effect estimates such that they become ``non-significant/not credible'', and it
is easy to show that the methods are under some circumstances also technically 
equivalent.
To illustrate this, we consider again the meta-analysis on the association
between corticosteroids and COVID-19 mortality \citep{REACT2020} which gave the
pooled log odds ratio estimate $\hat \theta = -0.42$ with standard error
$\sigma = 0.11$, posterior precision 
$\delta'=83.8$ and test statistic 
$z=\hat \theta/\sigma= -3.81$.

Using the \citet{Rosenberg2005} approach %%(based on normality instead of the $t$-distribution), 
(as implemented in the \texttt{fsn()} function from the \texttt{metafor} package)
we find that at least $N = 20$ additional but unpublished 
non-significant findings are needed to make the published meta-analysis effect
non-significant.
If instead, we challenge the overall estimate with AnCred, we obtain the
relative prior variance $g = 
0.36$ using equation \eqref{eq:sspv}, 
so $\tau^2 = 0.0043$. 
Taking into account the average precision
$\delta' / n = 
11.98$ of the different effect estimates estimates in the 
meta-analysis leads to ${N} = n/(\delta' \cdot \tau^2) = 19.5$ which 
is equivalent to the fail-safe $N$ result after rounding to the next larger
integer.

\subsection{Intrinsic credibility}
The Problem of Priors is at its most challenging in the context of entirely
novel ``out of the blue'' effects for which no obviously relevant external evidence
exist. By their nature, such findings often attract considerable interest both
within and beyond the research community, making their reliability of 
particular importance. Given the absence of external sources of evidence, 
\citet{matthews2018} proposed the concept of \emph{intrinsic credibility}. 
This requires that the evidential weight of an unprecedented finding is 
sufficient to put it in conflict with the sceptical prior rendering it 
non-credible. In the AnCred framework, this 
implies a finding possesses intrinsic credibility at level $\alpha$ if the 
estimate $\hat{\theta}$ is outside the corresponding sceptical prior interval
$[-S, S]$ extracted using Reverse-Bayes from the finding itself, 
\ie $\hat{\theta}^2 > S^2$ with $S$ given in \eqref{eq:S}.  
Matthews showed this implies an unprecedented finding 
is intrinsically credible at level $\alpha=0.05$ if its $p$-value does 
not exceed 0.013. 

\citet{held2019} refined the concept by suggesting the use of a prior-predictive
check \citep{box:1980, evans.moshonov2006} to assess potential prior-data
conflict. With this approach the uncertainty of the estimate $\hat{\theta}$
is also taken into account since it is based on the prior-predictive 
distribution, in this case 
$\hat\theta \sim \Nor(0, \sigma^2 + \tau^2=\sigma^2 \, (1+g))$ with $g$ 
as given in \eqref{eq:sspv}. 
Intrinsic credibility
is declared if the (two-sided) tail-probability 
$$p_{\text{Box}} = \P\left(\chi^2_1 \geq \hat \theta^2/(\sigma^2+\tau^2)\right) = \P\left(\chi^2_1 \geq z^2/(1+g)\right)$$ 
of $\hat{\theta}$ under the
prior-predictive distribution is smaller than $\alpha$.
It turns out that the $p$-value associated with $\theta$ needs to be at least as
small as 0.0056 to obtain intrinsic credibility at level $\alpha=0.05$, providing 
another principled argument for 
the recent proposition to lower the $p$-value threshold for the claims of new
discoveries to 0.005 \citep{Benjamin2017}. A simple check for intrinsic 
credibility is based on the \emph{credibility ratio}, the ratio of the upper to 
the lower limit (or vice versa) of a confidence interval for a significant
effect size estimate. If the credibility ratio is smaller than 5.8 
then the result is intrinsically credible \citep{held2019}. 
This holds for confidence intervals at all possible 
values of $\alpha$, not just for the 0.05 standard. 
For example, in the RECOVERY study the 95\% confidence interval for the
log-odds ratio ranges from $-0.82$ to 
$-0.25$, so the credibility ratio is
$-0.82/-0.25 = 
3.27 < 5.8$ and the result is intrinsically 
 credible at the standard 5\% level.

\subsubsection{Replication of effect direction}

Whether intrinsic credibility is assessed based on the
prior or the prior-predictive distribution, it depends on the level $\alpha$
in both cases. To remove this dependence, \citet{held2019} proposed to consider
the smallest level at which intrinsic credibility can be established,
defining the $p$-value for intrinsic credibility 
\begin{align*}
  \pIC
  &= 2\left\{1 - \Phi\left(\frac{\abs{z}}{\sqrt{2}}\right)\right\},
\end{align*}
see \citet[section 4]{held2019} for the derivation. Now $z=\hat \theta/\sigma$, so
compared to the standard
$p$-value $p=2\left\{1 - \Phi\left(\abs{z}\right)\right\}$, the
$p$-value for intrinsic credibility is based on twice the variance $\sigma^2$ of the estimate
$\hat{\theta}$. Although motivated from a different perspective, inference based on
intrinsic credibility thus mimics the \emph{doubling the variance rule} advocated
by \citet{CopasEguchi2005} as a simple means of adjusting for model uncertainty.

Moreover, \citet{held2019} showed that $\pIC$ is connected to $\prep$
\citep{Killeen2005}, the probability that a replication will result in
an effect estimate $\hat{\theta}_r$ in the same direction as the
observed effect estimate $\hat{\theta}$, by
$\prep = 1 - \pIC/2$.
% \citet{Killeen2005} proposed $p_{\text{rep}}$ as a more intuitive measure for
% the strength of evidence that an effect estimate carries.
Hence, an intrinsically credible estimate at a small level $\alpha$ 
will have high chance of replicating since
$\prep \geq 1 - \alpha/2$. Note that $\prep$ lies between 0.5 and 1 with
the extreme case $\prep=0.5$ if $\hat \theta=0$.

As an example, the $p$-value for intrinsic credibility for the RECOVERY
trial finding (with $p$-value $p=0.0002$) cited earlier is
$\pIC = 0.01$ and thus the probability of the replication
effect going in the same direction % replication leading to the same direction of effect 
(\ie reduced mortality in this case) is $0.995$. 
In contrast, the finding from
the smaller REMAP-CAP trial (with $p=0.29$) leads to $\pIC = 0.46$, and 
the probability of effect direction replication is hence only $0.77$.

\section{Reverse-Bayes methods with Bayes factors}
\label{sec:bfs}
The AnCred procedure as described above uses posterior credible
intervals as a means of quantifying evidence. However,
quantification of evidence with Bayes factors is a more principled solution
for hypothesis testing in the Bayesian framework \citep{jeffreys:1961, kass1995}.
Bayes factors enable direct probability
statements about null and alternative hypothesis and they can also quantify
evidence \emph{for} the null hypothesis, both are impossible with indirect 
measures of evidence such as $p$-values \citep{HeldOtt2018}.
Reverse-Bayes approaches combined with Bayes factor methodology was pioneered 
in \citet{Carlin1996} but then remained unexplored until 
\citet{Pawel2020b} proposed an extension of AnCred where Bayes factors
are used as a means of quantifying evidence. Rather than 
determining a prior such that a finding becomes ``non-credible''
in terms of a posterior credible interval, this approach determines a prior such
that the finding becomes ``non-compelling'' in terms of a Bayes factor.
In the second step of the procedure, the plausibility of this prior is 
quantified using external data from a replication study. Here, we will illustrate the 
methodology using only an original study; we mention extensions for replications 
in Section \ref{sec:extensions}.

\subsubsection*{Sceptical priors}\label{sec:sceptical}
A standard hypothesis test compares the null hypothesis 
$H_0\colon$ $\theta = 0$ to the alternative $H_1\colon$ $\theta \neq 0$. 
Bayesian hypothesis testing requires specification of a prior distribution of 
$\theta$ under $H_1$. A typical choice is a local 
alternative, a unimodal symmetric prior distribution centred around the null
value \citep{MR2830762}. We consider again the sceptical prior
$\theta \given H_1 \sim \Nor(0, \tau^2 = g \cdot \sigma^2)$ 
with relative prior variance $g$ for this purpose. 
This leads to the Bayes factor comparing $H_0$ to $H_1$ being
\begin{align*}
  \BF_{01} = 
  \sqrt{1 + g} \cdot
  \exp\left\{-\frac{g}{1 + g} \cdot \frac{z^2}{2} \right\}.
\end{align*}
Yet again, the amount of evidence which the data provide against the null
hypothesis depends on the prior parameter $g$; As $g$ becomes smaller
($g \downarrow 0$), the null and the alternative will become indistinguishable, so
the data are equally likely under both ($\BF_{01} \to 1$). On the other hand,
for increasingly diffuse priors ($g \to \infty$), the null hypothesis will 
always prevail ($\BF_{01} \to \infty$) due to the 
Jeffreys-Lindley paradox \citep{Robert2014}. In between, the $\BF_{01}$
reaches a minimum at $g = \max\left\{z^2 - 1, 0\right\}$ leading to
\begin{align}
  \text{minBF}_{01} = 
  \begin{cases}
    \abs{z} \cdot \exp\left\{-z^2/2\right\} \cdot \sqrt{e} 
    & \text{if} ~ \abs{z} > 1 \\
    1 & \text{else}
  \end{cases}
  \label{eq:minBFnorm}
\end{align}
which is an instance of a \emph{minimum Bayes factor}, 
the smallest possible Bayes factor within a class of alternative hypotheses, 
in this case zero-mean normal alternatives 
\citep{els:1963, bs:1987, sbb:2001, HeldOtt2018}.

Reporting of minimum Bayes factors is one attempt of solving the
problem of priors in Bayesian inference. However, this bound may be
rather small and the corresponding prior unrealistic.
In contrast, the Reverse-Bayes approach makes the choice of the prior explicit
by determining the relative prior variance parameter $g$ 
such that the finding is no longer compelling,
followed by assessing the plausibility of this prior.
To do so, one first fixes $\BF_{01} = \gamma$, where $\gamma$ is a cut-off 
above which the result is no longer convincing, for example $\gamma = 1/10$,
the level for strong evidence according to \citet{jeffreys:1961}. 
The sufficiently sceptical relative prior variance is then given by
\begin{align}
\label{ggamma}
  g &= 
  \begin{cases} 
    -\dfrac{z^2}{q} - 1
    & ~~ \text{if} ~ 
    -\dfrac{z^2}{q} \geq 1 \\ 
    \text{undefined} & ~~ \text{else}
  \end{cases} \\
  \text{where} ~ q &= 
  \text{W} \left(-\frac{z^2}{\gamma^2} \cdot
  \exp\left\{-z^2\right\}\right)
  \nonumber 
\end{align}
where $\text{W}(\cdot)$ is the Lambert W function \citep{Corless1996}, 
see \citet[Appendix B]{Pawel2020b} for a proof.

The sufficiently sceptical relative prior variance $g$ exists only for
a cut-off $\gamma$ if $\text{minBF}_{01} \leq \gamma$, similar to standard
AnCred where it exists only at level $\alpha$ if the original finding was 
significant at the same level. In contrast to standard AnCred, however, if the
sufficiently sceptical relative prior variance $g$ exists, there are always two
solutions,
% that can
% be obtained through the two branches of the Lambert $W$-function 
% (see \citet{Pawel2020b} for details), 
a consequence of the Jeffreys-Lindley paradox: 
%\citep{Robert2014}: For increasingly diffuse priors,
% the null hypothesis will always prevail, \ie $\BF_{01} \to \infty$ 
% as $g \to \infty$. 
If $\BF_{01}$ decreases in $g$ below the chosen cut-off 
$\gamma$, after attaining its minimum %at $g = \max\{t^2 - 1, 0\}$
it will monotonically increase and intersect
a second time with $\gamma$, admitting a second solution for the sufficiently
sceptical prior.

\begin{figure}[!htb]
\begin{knitrout}
\definecolor{shadecolor}{rgb}{0.969, 0.969, 0.969}\color{fgcolor}

{\centering \includegraphics[width=\linewidth]{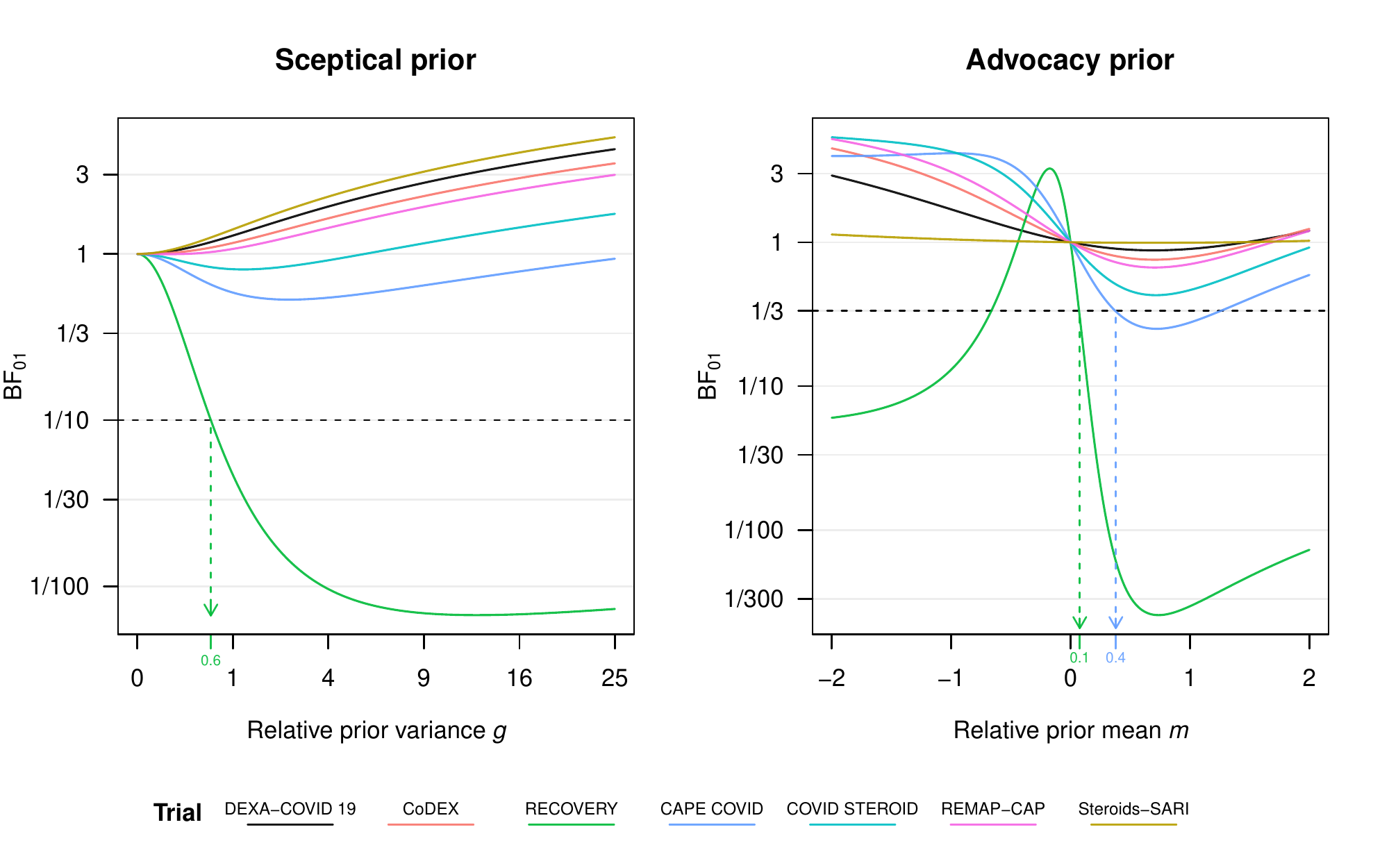} 

}

\end{knitrout}
\caption{Illustration of the AnCred with Bayes factors procedure using
  the findings from the meta-analysis on the association of COVID-19
  mortality and corticosteroids.  The left plot shows the Bayes factor
  $\BF_{01}$ as a function of the relative variance $g$ of the
  sceptical prior.  The result from the RECOVERY trial is challenged
  with a sceptical prior such that $\BF_{01} = 1/10$, for the other
  trials such a prior does not exist.  The right plot shows the Bayes
  factor $\BF_{01}$ as a function of the relative mean
  $m = \mu/\hat{\theta}$ of the advocacy prior where the coefficient
  of variation from the prior is fixed to
  $\text{CV} = \tau/\mu = 1/z(\gamma=1/3) = 0.67$,
  where $z(\gamma)$ is given in \eqref{eq:zgamma}.  The RECOVERY and the
  CAPE COVID findings are challenged such that $\BF_{01} = 1/3$, for
  the other trials such a prior does not exist.}
\label{fig:bf}
\end{figure}

We now revisit the meta-analysis example considered earlier: The left plot in Figure 
\ref{fig:bf} shows the Bayes factor $\BF_{01}$ as a function of the relative prior 
variance $g$ for each finding included in the meta-analysis.
Most of them did not include a great number of participants
and thus provide little evidence against the null for any value of 
the relative prior variance $g$. 
In contrast, the finding from the RECOVERY trial \citep{RECOVERY2020} 
provides more compelling evidence and can be challenged up to 
$\text{minBF}_{01} = 1/148.9$. 
For example, we see in Figure \ref{fig:bf} that the sceptical 
prior variance needs to be $g = 0.59$, so
$1.69$ times
smaller than the variance of the effect estimate, such that the finding
is no longer compelling at level $\gamma = 1/10$. 
This translates to a 95\% prior credible interval from
0.8 to 1.24 for the OR. Hence,
a sceptic might still consider the RECOVERY finding to be
unconvincing, despite its minimum BF being very compelling, if external 
evidence supports ORs in that range.
% of paste(round(priorCI, 2), collapse = " to ") with 95\% probability.
% Recall that for the same finding we needed to choose a sceptical prior 
% variance round(sigma2ex1/tau2ex1, 1) times smaller than the variance
% of the estimate to render its posterior ``non-credible'' at level 
% $\alpha = 0.05$ (Figure \ref{fig:anCredEx}).
Note that also
$g^\prime = 8190$ gives a Bayes factor of 
 $\BF_{01} = 1/10$, however, such a large relative prior variance 
represents ignorance rather than scepticism and is less useful for
Reverse-Bayes inference.

The plausibility of the sufficiently sceptical prior can be evaluated in light
of external evidence, but what should we do in the absence of such?
We could again use the \citet{box:1980} prior-predictive check, however, the 
resulting tail probability is difficult to compare to the Bayes-factor cut-off 
$\gamma$. When a specific alternative model to the null is in mind, Box also 
suggested to use a Bayes factor contrasting the two models. %as a diagnostic check. 
Following this approach, \citet{Pawel2020b} proposed to define a second Bayes 
factor contrasting the 
sufficiently sceptical prior to an optimistic prior, 
which they defined as $\theta \given H_2 \sim \Nor(\hat{\theta}, \sigma^2)$ 
the posterior of $\theta$ based on the data and the reference prior 
$f(\theta) \propto 1$. 
%, or even $\theta = \hat{\theta}$.
% \todo{should scepical BF be moved in the Discussion just as sceptical $p$?}
% \citet{Pawel2020b} evaluated this Bayes factor on external data from a replication
% study, however, it is also possible to evaluate it on the observed effect estimate. 
We can then conclude that the effect estimate is intrinsically credible at level 
$\gamma$ if the data favour the optimistic prior over the sufficiently sceptical 
prior at a higher level than $1/\gamma$ (\ie if $\BF_{12} \leq \gamma$), analogously 
to intrinsic credibility based on significance.
For example, we obtain $\BF_{12} = 1/64$ for the finding
from the RECOVERY trial, so it is intrinsically credible at $\gamma = 1/10$. 
To remove the dependence on the choice of 
$\gamma$, one can then determine the smallest cut-off $\gamma$ where intrinsic 
credibility can be established, defining a Bayes factor for intrinsic 
credibility similar to the definition of the $p$-value for intrinsic credibility.
For the RECOVERY finding, this turns out to be 
$\BF_{\scriptsize \mbox{IC}} = 1/25$.

\subsubsection*{Advocacy priors}
A natural question is whether we can also define an advocacy prior,
a prior which renders an uncompelling finding compelling, 
in the AnCred framework with Bayes factors.
In traditional AnCred, advocacy priors always exist %with significance 
since one can always find a prior that, when combined with 
the data, can overrule them. This is fundamentally different to inference 
based on Bayes factors, where the prior is not synthesized with the data, but 
rather used to predict them.
A classical result due to 
\citet{els:1963} states that if we consider the class of all possible priors
under $H_1$, the minimum Bayes factor is given by 
\begin{equation}\label{eq:ELSminBF}
  \text{minBF}_{01} = \exp\left\{-z^2/2\right\}
\end{equation}
which is obtained for 
$H_1$: $\theta = \hat{\theta}$. This implies that a non-compelling finding
can not be ``rescued'' further than to this bound. 
For example, for the finding from the REMAP-CAP trial \citep{REMAPCAP2020} the
bound is unsatisfactorily 
$\text{minBF}_{01} = 1/1.7$, 
so at most ``worth a bare mention'' according to \citet{jeffreys:1961}.

Putting these considerations aside, we may still consider the class of 
$\Nor(\mu, \tau^2)$ priors under the alternative $H_1$. 
The Bayes factor contrasting $H_0$ to $H_1$ is then given by
\begin{align*}
  \BF_{01} 
  &= \sqrt{1 + \tau^2/\sigma^2} \cdot \exp\left\{-\frac{1}{2}\left[
  \frac{\hat{\theta}^2}{\sigma^2} - \frac{(\hat{\theta} - \mu)^2}{\sigma^2 + 
  \tau^2}\right]\right\}. 
  % &= \sqrt{1 + g} \cdot \exp\left\{-\frac{z^2}{2}\left[1 - 
  % \frac{(1 - m)^2}{1 + g}\right]\right\}.
\end{align*}
The reverse-Bayes approach now 
determines the prior mean $\mu$ and variance $\tau^2$ which lead to the Bayes 
factor $\BF_{01}$ being just at some cut-off $\gamma$. However, if both 
parameters are free, there are infinitely many solutions to 
$\BF_{01} = \gamma$, if any exist at all. 
The traditional AnCred framework resolves this by restricting the class of possible
priors to advocacy priors with fixed coefficient of variation of
$\text{CV} = \tau/\mu = 1/z_{\alpha/2}$. 
We can translate this idea to the Bayes factor AnCred framework and fix 
the prior's coefficient of variation to $\text{CV} = 1/z(\gamma)$, 
where %% $z(\gamma)$ is a $z$-value corresponding to $\text{minBF}_{01} = \gamma$. 
%% Inverting equation %\eqref{eq:minBFnorm} 
%% \eqref{eq:ELSminBF} 
%% leads to 
\begin{align}
  % \abs{z(\gamma)} = 
  % \begin{cases}
  %   \sqrt{-\text{W}_{\scriptscriptstyle{-1}}\left(-\gamma^2 / e \right)} 
  %   & ~ \text{if} ~ \gamma < 1 \\
  %   \text{undefined} & ~ \text{else}
  % \end{cases}
  z(\gamma) = \sqrt{- 2 \, \log \gamma}, \label{eq:zgamma}
\end{align}
obtained by solving \eqref{eq:ELSminBF} for $z$
 with $\text{minBF}_{01} = \gamma$. 
% with $\text{W}_{\scriptscriptstyle{-1}}(\cdot)$ the \todo{second?} branch of the Lambert W
% function \todo{drop details?} satisfying $\text{W}(x) \leq -1$ for $-1/e \leq x < 0$.
The advocacy prior thus carries the same evidential weight as data with
$\text{minBF}_{01} = \gamma$. Moreover, the determination of the 
prior parameters becomes more feasible since there is only one free parameter
left (either $\mu$ or $\tau^2$).

The right plot in Figure \ref{fig:bf} illustrates application of the procedure 
on data from the meta-analysis on association between COVID-19 mortality and 
corticosteroids. 
The coefficient of variation of the advocacy prior is fixed to 
$\text{CV} = 1/z(\gamma=1/3) = 0.67$ and thus the Bayes factor 
$\BF_{01}$ only depends on the relative mean $m = \mu/\hat{\theta}$. 
Under the 
sceptical prior only the RECOVERY finding could be challenged at 
$\gamma = 1/3$ (where $z(\gamma)=1.5$ corresponds to $\alpha=13$\%). With the advocacy prior this is now also possible for the CAPE COVID 
finding \citep{Dequin2020}, where a prior with mean 
$\mu = m \cdot \hat{\theta} = 0.37 \cdot 
(-0.79) = 
-0.29$ and standard deviation 
$\tau = \text{CV} \cdot \mu
= 0.2$
is able to make the finding compelling at $\gamma = 1/3$. %% This corresponds
The corresponding prior %% to a 87\% 
credible interval for the OR at level $1-\alpha$ ranges from
0.55 to 1, so advocates may
still consider the ``non-compelling'' finding as providing
moderate evidence in favour of a benefit, if external evidence supports 
mortality reductions in that range.
% (in terms of odds) between
% paste(round(CapeAdvCIOR, 2), collapse = " and ") with 95\% probability. 
Note that the advocacy prior may not be unique, \eg for the 
CAPE COVID finding the prior with relative mean
$m^\prime = 1.26$ and standard deviation
$\tau^\prime = 0.67$ 
% (95\% prior credible interval from 
% paste(round(CapeAdv2CIOR, 2), collapse = " to ") on OR scale)
also renders the data as just compelling at $\gamma = 1/3$. We recommend to
choose the prior with $m$ closer to zero, as it is the more conservative choice. 

\section{Reverse-Bayes Analysis of the False Positive Risk}
\label{sec:p.equals}
Application of the Analysis of Credibility with Bayes factors as
described in Section \ref{sec:bfs}
%\eqref{sec:bfs} 
assumes some familiarity with
Bayes factors as measures of evidence.  \citet{Colquhoun:2019} argued
that very few nonprofessional users of statistics are familiar with
the notion of Bayes factors or likelihood ratios. He proposes to
quantify evidence with the {\em false positive risk}, ``if only
because that is what most users still think, mistakenly, that that is
what the $p$-value tells them''. More specifically,
\citet{Colquhoun:2019} defines the \ac{FPR} as the posterior
probability that the point null hypothesis $H_0$ of no effect is true
given the observed $p$-value $p$, \ie
$\FPR = \Pr(H_0 \given p)$. As before, $H_0$ corresponds to the point
null hypothesis $H_0\colon$ $\theta = 0$. Note also that we take the
exact (two-sided) $p$-value $p$ as the observed ``data'', regardless
of whether or not it is significant at some pre-specified level, the
so-called ``$p$-equals'' interpretation of NHST \citep{Colquhoun:2017}.

$\FPR$ can be calculated based on the Bayes factor associated with $p$.
%% Calibrating $p$-values as posterior probabilities or Bayes factors
%% has a long history, going back to at least the 1960s \citep{els:1963},
%% see \citet{HeldOtt2018} for a recent review. 
%% These calibrations can be transformed to the $\FPR$ scale
%% by using Bayes' theorem.
% which makes them easier to interpret.
% The $\FPR$ can be calculated using Bayes' theorem, which requires
% specification of the prior probability $\Pr(H_1) = 1 - \Pr(H_0)$.
% Formulated in terms of odds, this theorem then reads:
For ease of presentation we invert Bayes' theorem \eqref{eq:eq0}  and obtain
\begin{equation}\label{eq:bayes}
      % \frac{1-\FPR}{\FPR} = \frac{\Pr(H_1 \given p)}{\Pr(H_0 \given p)}
      % = {\mbox{BF}_{10}} \, \frac{\Pr(H_1)}{\Pr(H_0)},
      \frac{\FPR}{1-\FPR} = \frac{\Pr(H_0 \given p)}{\Pr(H_1 \given p)}
      = {\mbox{BF}_{01}} \, \frac{\Pr(H_0)}{\Pr(H_1)},
\end{equation}
% where $\mbox{BF}_{10}=1/\mbox{BF}_{01}$ is the Bayes factor for $H_1$ against $H_0$,
% computed from the observed $p$-value $p$.
where $\mbox{BF}_{01}=1/\mbox{BF}_{10}$ is the Bayes factor for $H_0$ against $H_1$,
computed directly from the observed $p$-value $p$.

%% Given any two of the following three quantities,
%% the prior probability $\Pr(H_0)$, % $\Pr(H_1)$, 
%% % the Bayes factor $\mbox{BF}_{10}$ (or $\mbox{BF}_{01}$)
%% the Bayes factor $\mbox{BF}_{01}$ (or $\mbox{BF}_{10}$)
%% and the $\FPR$,
%% the third quantity can be computed using Equation \eqref{eq:bayes}.\todo{MO: drop this?}
%% % The most common approach, called ``forward-Bayes'' here,
The common 'forward-Bayes' approach is to compute the $\FPR$ from the
prior probability $\Pr(H_0)$ %$\Pr(H_1)$
and the Bayes factor with \eqref{eq:bayes}.  However, the prior
probability $\Pr(H_0)$ %$\Pr(H_1)$
is usually unknown in practice and often hard to assess.  
This can be resolved via the Reverse-Bayes approach
\citep{Colquhoun:2017,Colquhoun:2019}: Given a $p$-value and a false
positive risk value, calculate the corresponding prior probability
$\Pr(H_0)$ %$\Pr(H_1)$
that is needed to achieve that false positive risk.  Of specific
interest is the value FPR = 5\%, because many scientists believe that a
Type-I error of 5\% is equivalent to a FPR of 5\% \citep{Greenland2016}.
This is of course not true and we follow \citet[Example 1]{bs:1987}
and use the reverse-Bayes approach
to derive the necessary prior assumptions on $\Pr(H_0)$ to achieve
FPR = 5\% with Equation \eqref{eq:bayes}: 
\begin{equation}\label{eq:prior.prob}
     % \Pr(H_1) = \left[ 1 + \frac{\FPR}{(1-\FPR)\mbox{BF}_{01}}  \right]^{-1}.
     \Pr(H_0) = \left[ 1 + \frac{1-\FPR}{\FPR} \cdot \mbox{BF}_{01}  \right]^{-1}.
\end{equation}

\citet[appendix A.2]{Colquhoun:2017} uses a Bayes factor based on the
$t$-test, but for compatibility with the previous sections we assume normality
of the 
underlying test statistic. We consider Bayes factors under all simple
alternatives, but also Bayes factors under local normal priors, see
\citet{HeldOtt2018} for a detailed comparison. 

Instead of working with a Bayes factor for a specific prior distribution, 
we prefer to work with the minimum Bayes factor
$\mbox{minBF}_{01}$ as introduced in Section \ref{sec:sceptical}.
%% The minimum Bayes factor is a lower
%% bound on the Bayes factor $\mbox{BF}_{01}$ over a specific class of
%% alternatives. %% Minimum Bayes factors for several
%% common hypothesis tests such as the $t$-, $z$-, $\chi^2$- and $F$-test
%% are implemented in the R package \texttt{pCalibrate}
%% \citep{pCalibrate2020} available on the Comprehensive R Archive
%% Network (CRAN).  Minimum Bayes factors can also be incorporated into
%% false positive risk analyses.  
In what follows we will use the minimum Bayes factor based on the
$z$-test \citep[Section 2.1 and 2.2]{HeldOtt2018}. %%\todo{MO: I would also mention Section 2.2 (of \citet{HeldOtt2018}) on local alternatives here since the minBF (10) is introduced there}.  
The minimum Bayes factor based on the
$z$-test among all possible priors can be computed using the function
\texttt{zCalibrate} in the package \texttt{pCalibrate}.   
%% with the option
%% \texttt{alternative = "simple"} (and \texttt{type = "two.sided"},
%% which is the default). 
The option \texttt{alternative = "local"}  gives the
minBF \eqref{eq:minBFnorm} under local normal priors.

Let $\mbox{minBF}_{01}$
denote the minimum Bayes factor over a specific class of alternatives.
From equation \eqref{eq:prior.prob} we obtain the inequality
\begin{equation}\label{eq:prior.prob2}
     % \Pr(H_1) = \left[ 1 + \frac{\FPR}{(1-\FPR)\mbox{BF}_{01}}  \right]^{-1}.
     \Pr(H_0) \leq \left[ 1 + \frac{1-\FPR}{\FPR} \cdot \mbox{minBF}_{01}  \right]^{-1}.
\end{equation}
The right-hand side is thus an upper bound on the prior probability
$\Pr(H_0)$ for a given $p$-value to achieve a pre-specified FPR value.

There are also minBFs not based on the $z$-test statistic, but
directly on the (two-sided) $p$-value $p$, the so-called
``$- e \, p \log p$'' \citep{sbb:2001} calibration
\begin{equation}
\label{eq:sbb}
{\mbox{minBF}} =  \left\{ \begin{array}{ll} - e \, p \log p & \mbox{ for } p < 1/e \\
  1 & \mbox{ otherwise, } \end{array}   \right.
\end{equation}
and the ``$- e \, q \log q$'' calibration, where  $q = 1-p$
\citep[Section 2.3]{HeldOtt2018}: 
\begin{equation}
\label{eq:sbb2}
{\mbox{minBF}} =  \left\{ \begin{array}{ll} - e \, (1-p) \log (1-p) & \mbox{ for } p < 1-1/e \\
  1 & \mbox{ otherwise. } \end{array}   \right.
\end{equation}
For small $p$,  equation \eqref{eq:sbb2} can be simplified to
${\mbox{minBF}} \approx e \, p$, which mimics the \citet{Good1958} transformation of
$p$-values to Bayes factors \citep{Held5855}.

The two $p$-based calibrations are also available
in the package \texttt{pCalibrate}. 
They carry less assumptions than the minimum Bayes factors based on
the $z$-test under normality. The ``$- e \, p \log p$'' provides a
general bound under all unimodal and symmetrical local priors for $p$-values 
from $z$-tests \citep[Section 3.2]{sbb:2001}.
The ``$- e \, q \log q$'' calibration is more conservative and
gives a smaller bound on the Bayes factor than the ``$- e \, p \log p$'' calibration. 
It can be viewed as a general lower bound under simple alternatives where 
the direction of the effect is taken into account, see 
\citet[Section 2.1 and 2.3]{HeldOtt2018}. %%\todo{MO: also mention Section 2.3 (of \citet{HeldOtt2018}), where this relationship is described?}

\begin{figure}[!htb]
\begin{knitrout}
\definecolor{shadecolor}{rgb}{0.969, 0.969, 0.969}\color{fgcolor}

{\centering \includegraphics[width=\linewidth]{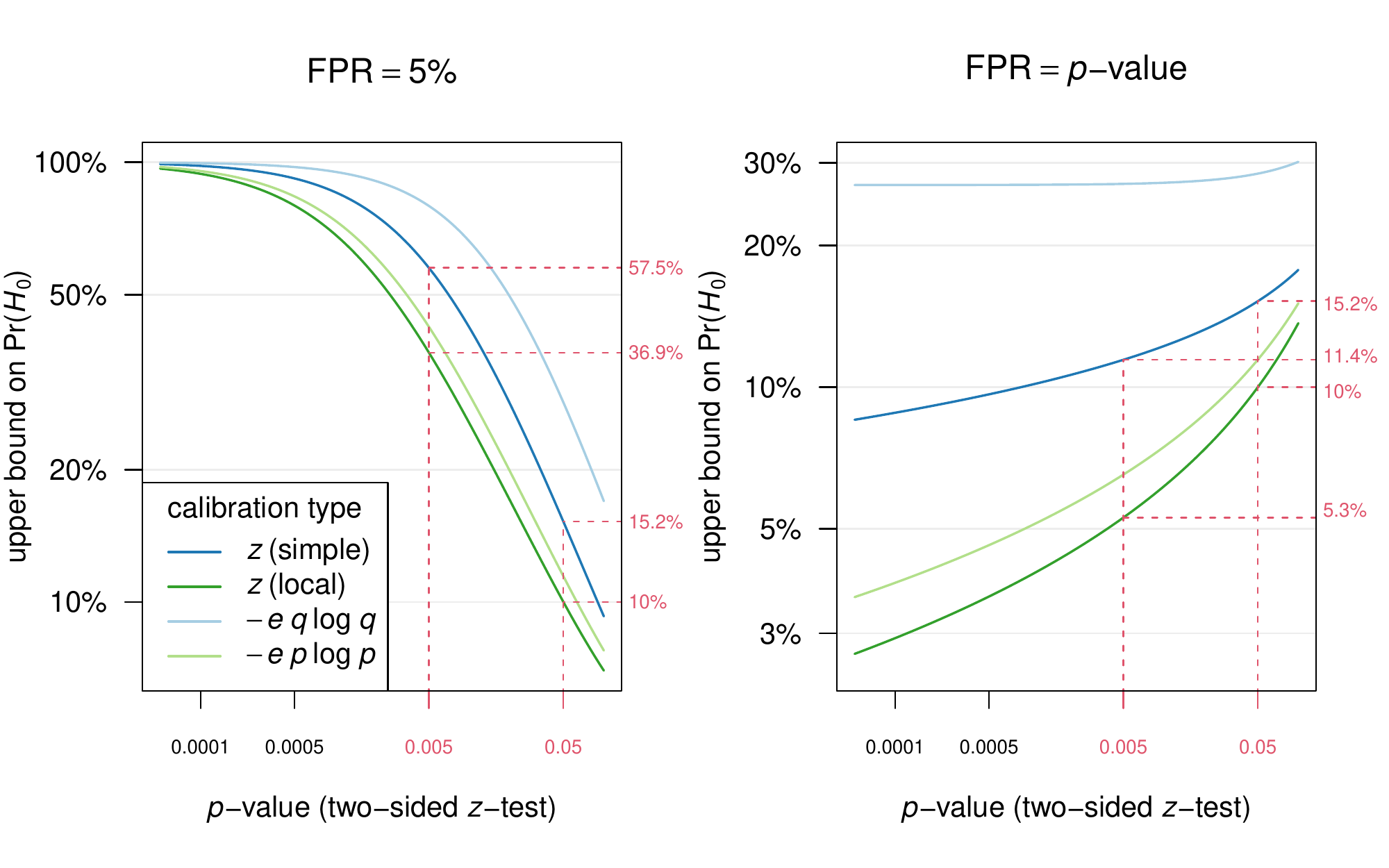} 

}

\end{knitrout}
\caption{The left plot shows the upper bound on the prior probability $\Pr(H_0)$
to achieve a false positive risk of 5\% as a function of the $p$-value calibrated
with either a $z$-test calibration (simple and local alternatives) or with the 
``$- e \, p \log p$'' or ``$- e \, q \log q$'' calibrations, respectively.
The right plot shows the upper bound on $\Pr(H_0)$ as a function of the
$p$-value using the same calibrations but assuming the $p$-value equals the $\FPR$.}
\label{fig:fpr}
\end{figure}

The left plot in Figure \ref{fig:fpr} shows the resulting upper bound
on the prior probability $\Pr(H_0)$ as a function of the two-sided
$p$-value if the FPR is fixed at 5\%. 
For $p=0.05$, the
``$- e \, p \log p$'' bound is around
11\% and
28\% for
the ``$- e \, q \log q$'' calibration.  The corresponding values based
on the $z$-test are slightly smaller
(10\% and
15\%,
respectively). All the probabilities are below the 50\% value of
equipoise, illustrating that borderline significant result with
$p \approx 0.05$ do not provide sufficient evidence to justify an FPR
value of 5\%. For $p=0.005$, the upper bounds are closer to 50\% 
(37\% for local and
57\%
for simple alternatives).

Turning again to the example from the RECOVERY trial
\citep{RECOVERY2020}, the $p$-value associated with the estimated
treatment effect is $p = 0.0002$. The left plot
in Figure \ref{fig:fpr} shows that the false positive risk can safely
be assumed to be around 5\% (or lower), since the upper bound on
$\Pr(H_0)$ are all very large for such a small $p$-value. %% However, to
%% justify the value $\FPR = round(100*pRECOVERY, 2)$\% the prior
%% probability on $H_0$ must be quite small under all calibrations, see
%% the right plot in Figure \ref{fig:fpr}. 

Fixing FPR at the 5\% level may be considered as arbitrary. 
Another widespread misconception is the belief that 
that the FPR is equal to the $p$-value. 
%% Similar to the Reverse-Bayes analysis for the misconception ``$\FPRP = \alpha$''
%%  studied in Section \ref{sec:p.less}, 
\citet{held2013} used a reverse-Bayes approach to investigate which
prior assumptions are required such that $\FPR=p$ holds.  
Combining \eqref{eq:prior.prob} with the ``$- e \, p \log p$'' calibration \eqref{eq:sbb} gives
the explicit condition
\[
  \Pr(H_0) \leq 1/\left\{ 1 - e \, (1-p) \, \log(p) \right\}
\]
whereas the ``$- e \, q \log q$'' calibration \eqref{eq:sbb2} leads to 
\[
  \Pr(H_0) \leq 1/\left\{ 1 - e \, \frac{(1-p)^2}{p} \, \log(1-p) \right\} \approx 1/\left\{ 1 + e \, (1-p) \right\},
\]
which is approximately $1/(1+e)=26.9$\% for small $p$. 

The right plot in Figure \ref{fig:fpr} compares the bounds based on
these two calibrations with the ones obtained from simple respectively
local alternatives. We can see that strong assumptions on $\Pr(H_0)$
are needed to justify the claim $\FPR=p$: $\Pr(H_0)$ cannot be larger
than 15.2\% if the $p$-value is conventionally significant
($p<0.05$). For $p<0.005$, the bound drops further to 11.4\%.  Even
under the conservative ``$- e \, q \log q$'' calibration, the upper
bound on $\Pr(H_0)$ is $26.9$\% for small
$p$ and increases only slightly for larger values of $p$.  This
illustrates that the misinterpretation $\FPR=p$ only holds if the
prior probability of $H_0$ is substantially smaller than 50\%, an
assumption which is questionable in the absence of strong external
knowledge.

\section{Discussion}

\subsection{Extensions, work in progress and outlook}
\label{sec:extensions}
The Reverse-Bayes methods described above have focused on the comparison of the 
prior needed for credibility with findings from other studies and/or more general
insights. However, replication studies make an obvious additional source of 
external evidence, as these are typically conducted to confirm original findings
by repeating their experiments as closely as possible. The question is then 
whether the original findings have been successfully ``replicated'', currently of 
considerable concern to the research community.   To date, there remains
no consensus on the precise meaning of replication in a statistical sense.
The proposal of \citet{Held2020} \citep[see also][]{held_etal2020} was
to challenge the original finding using AnCred, as described in Section 
\ref{sec:AnCred}, and then evaluate the plausibility of the resulting
prior using a prior-predictive check on the data from a replication 
study. A similar procedure but using AnCred based on Bayes factors as in
Section \ref{sec:bfs} was proposed in \citet{Pawel2020b}.
Reverse-Bayes inference seems to fit naturally into this setting as it provides 
a formal framework to challenge and substantiate scientific findings.

Apart from using data from a replication study, there are also other
possible extensions of AnCred: We proposed either prior-predictive
checks \citep{box:1980, evans.moshonov2006} or Bayes-factors
\citep{jeffreys:1961, kass1995} for the formal evaluation of the
plausibility of the priors derived through Reverse-Bayes. Other
methods could be used for this purpose, for example, Bayesian measures
of surprise \citep{Bayarri2003}.  Furthermore, AnCred in its current
state is derived assuming a normal likelihood for the effect estimate
$\hat{\theta}$. This is the same framework as in standard
meta-analysis and provides a good approximation for studies with
reasonable sample size \citep{Carlin1992}. For the comparison of
binomial outcomes with small counts, the normal approximation of the
log odds ratio could be improved with a Yates continuity correction
\citep[Sec.~2.4.1]{spiegelhalter2004} or replaced with the exact
profile likelihood of the log odds ratio
\citep[Sec.~5.3]{Held.SabanesBove2020}.  Likewise, more robust prior
distributions could be considered such as double-exponential or
Student $t$-distributions \citep{PericchiSmith1992}. For example, 
\citet{Fuquene_etal2009} investigate the use of robust priors in an
application to binomial data from a randomized clinical trial.

\subsection{Conclusions}
The inferential advantages of Bayesian methods are increasingly recognised within 
the statistical community. However, among the majority of working researchers 
they have failed to make any serious headway, and retain a reputation for complex
and ``controversial''. 
%% In this review, 
We have outlined how an idea that began with Jack Good's proposal
for resolving the ``Problem of priors'' over 70 years ago \citep{good:1950} has
experienced a renaissance over recent years.
The basic idea is to invert Bayes' theorem: a specified posterior is combined 
with the data to obtain the Reverse-Bayes prior, which is then used for further 
inference. 
This approach is useful in situations where it is difficult to decide what constitutes
a reasonable prior, but easy to specify the posterior which would lead to a 
particular decision. A subsequent prior-to-data conversion \citep{greenland:2006} 
helps to assess the weight of the Reverse-Bayes prior in relation to the actual data.

We have shown that the Reverse-Bayes methodology is 
useful to extract more insights from the results typically reported in a
meta-analysis. It facilitates the computation of 
prior-predictive checks for conflict diagnostics \citep{Presanis2013} 
and 
%% Starting with the work of \citet{matthews:2001,matthews:2001b}, the Reverse-Bayes 
%% methodology 
has been shown capable of addressing many common inferential 
challenges, including assessing the credibility of scientific findings
\citep{spiegelhalter2004,greenland:2011}, making sense
of ``out of the blue'' discoveries with no prior support
\citep{matthews2018, held2019}, estimating the probability 
of successful replications \citep{held2019, Held2020}, and extracting more 
insight from standard $p$-values while reducing the risk of misinterpretation
\citep{held2013,Colquhoun:2017,Colquhoun:2019}.
The appeal of Reverse-Bayes techniques has recently been widened by the 
development of inferential methods using both posterior probabilities and 
Bayes Factors \citep{Carlin1996, Pawel2020b}. 

These developments come at a crucial time for the role of statistical methods in 
research. Despite the many serious -- and now well-publicised – inadequacies of
NHST \citep{Wasserstein2016}, the research community has shown itself to be 
remarkably reluctant to abandon NHST. Techniques based on the Reverse-Bayes 
methodology of the kind described in this review could encourage the wider use
of Bayesian inference by researchers. As such, we believe they can play a key 
role in the scientific enterprise of the 21\textsuperscript{th} century.

\section*{Data and Software}
All analyses were performed in the R programming language version 
3.6.3 \citep{R}. 
Data and code to reproduce all analyses is available at 
\url{https://gitlab.uzh.ch/samuel.pawel/Reverse-Bayes-Code}.

\section*{Acknowledgments}
Support by the Swiss National Science Foundation (Project \# 189295)
is gratefully acknowledged. We are grateful to Sander Greenland for
helpful comments on a previous version of this article.

%% ** Highlights **

\section*{Highlights}

\begin{itemize}
\item \emph{What is already known?} \\
  Standard methods of statistical inference have led to a crisis in
  the interpretation of research findings. The adoption of standard
  Bayesian methods is hampered by the necessary specification of a
  prior level of belief.
\item \emph{What is new?} \\
  Reverse-Bayes methods open up new inferential tools of practical
  value for evidence assessment and research synthesis. 
\item \emph{Potential impact for RSM readers} \\
Reverse-Bayes methodology enables researchers 
to extract new insights from summary measures,
to assess the credibility of scientific findings and to reduce the risk of misinterpretation. 
\end{itemize}

%% ** The bibliograhy **
\bibliographystyle{ba}
%\bibliography{<bib-data-file>}% place <bib-data-file> in ./bib folder 
\bibliography{antritt}

%%%%%%%%%%%%%%%%%%%%%%%%%%%%%%%%%%%%%%%%%%%%%% 
%% Supplementary Material, if any, should   %%
%% be provided in {supplement} environment  %%
%% with title and short description.        %%
%%%%%%%%%%%%%%%%%%%%%%%%%%%%%%%%%%%%%%%%%%%%%%
% \begin{supplement}
% \stitle{Proof of equation \eqref{eq:tau2}}\label{app:app1}
% \sdescription{Proof of equation \eqref{eq:tau2}}
\begin{appendices}
\label{sec:appendices}
\section{Proof of equation \eqref{eq:mu}}
\label{app:app1}
Suppose that the estimate $\hat \theta$ is not significant
at level $\alpha$, so $z^2/z_{\alpha/2}^2<1$.  With
$U,L = \hat \theta \pm z_{\alpha/2} \, \sigma$ we have
$U + L = 2 \, \hat \theta$,
$U L = \hat \theta^2 - z_{\alpha/2}^2 \, \sigma^2$ and $U-L
= %% \sign{(U-L)} \,
2 \, z_{\alpha/2} \, \sigma$.

We therefore obtain with \eqref{eq:AL}:
\begin{equation*}
\mu = \frac{\mbox{AL}}{2} =  - \frac{2 \, \hat \theta}{2 \, (\hat \theta^2 - z_{\alpha/2}^2 \, \sigma^2)} \frac{(2 \, z_{\alpha/2} \, \sigma)^2}{2}
 =  \frac{2 \, \hat \theta \, z_{\alpha/2}^2 \, \sigma^2}{z_{\alpha/2}^2 \, \sigma^2 - \hat \theta^2}
 =  \frac{2 \, \hat \theta}{1 - z^2/z_{\alpha/2}^2}. 
\end{equation*}
The advocacy standard deviation is 
  $\tau = {\mbox{AL}}/{(2 \, z_{\alpha/2})} = {\mu}/{z_{\alpha/2}}$ and the
  coefficient of variation is therefore $\mbox{CV} = \tau/\mu = z_{\alpha/2}^{-1}$.

\end{appendices}

% ** Acknowledgements **
% \begin{acknowledgement}
% \end{acknowledgement}

\end{document}